# Proton magnetic resonance imaging with a nitrogen-vacancy spin sensor

D. Rugar[1]*, H. J. Mamin[1], M. H. Sherwood[1], M. Kim[1,2], C. T. Rettner[1], K. Ohno[3] and D. D. Awschalom[3,4]

[1]IBM Research Division, Almaden Research Center, San Jose, CA 95120, USA

[2]Center for Probing the Nanoscale, Stanford University, Stanford, CA 94305, USA

[3]Center for Spintronics and Quantum Computation, University of California, Santa Barbara, CA 93106, USA

[4]Institute for Molecular Engineering, University of Chicago, IL 60637, USA

*For correspondence: rugar@us.ibm.com

6/8/2014 1:07 PM

# Abstract

Nuclear magnetic resonance (NMR) imaging with nanometer resolution requires new detection techniques with sensitivity well beyond the capability of conventional inductive detection. Here, we demonstrate two dimensional imaging of $^1$H NMR from an organic test sample using a single nitrogen-vacancy center in diamond as the sensor. The NV center detects the oscillating magnetic field from precessing protons in the sample as the sample is scanned past the NV center. A spatial resolution of 12 nm is shown, limited primarily by the scan accuracy. With further development, NV-detected magnetic resonance imaging could lead to a new tool for three-dimensional imaging of complex nanostructures, including biomolecules.

Magnetic resonance imaging (MRI), with its ability to provide three-dimensional, elementally selective imaging without radiation damage, has had revolutionary impact in many fields, especially medicine and the neurosciences. Although challenging, extension of MRI to the nanometer scale could provide a powerful new tool for the nanosciences, especially if it can provide a means for non-destructively visualizing the full three-dimensional morphology of complex nanostructures, including biomolecules[1]. To achieve this potential, innovative new detection strategies are required to overcome the severe sensitivity limitations of conventional inductive detection techniques[2]. For example, by sensing attonewton magnetic forces, magnetic resonance force microscopy[3,4] (MRFM) has demonstrated three-dimensional imaging of proton nuclear magnetic resonance (NMR) in a biological sample with resolution on the order of 10 nm, but with the requirement of operating at cryogenic temperatures[5].

Nitrogen-vacancy (NV) centers in diamond offer an alternative detection strategy for nanoscale MRI that is operable at room temperature[6]. With their exceptional quantum mechanical properties, including long spin coherence time and convenient optical spin-state readout, individual NV centers can serve as atomic-size magnetometers with nanotesla magnetic field sensitivity[7–10]. Recently, NV centers have been used to sensitively detect both electron spin resonance and NMR for spins located either internal or external to the host diamond crystal[11–21]. In the case of electron spin resonance, imaging has been achieved with nanometer resolution and single spin sensitivity[22,23]. Progressing to nuclear spin imaging poses additional detection challenges since nuclear spins have at least 600 times smaller magnetic moment compared to electron spins. Here, we describe two-dimensional $^1$H NMR imaging of an organic test sample with lateral resolution on the order of 12 nm. With further development, the techniques presented here could provide a foundation for extending magnetic resonance imaging into important new application regimes, such as molecular structure imaging.

## Experiment fundamentals

The imaging experiment uses a scanning sample configuration where a near-surface NV center detects the magnetic field emanating from protons in the sample as the sample is mechanically scanned with nanometer precision (Fig. 1a). The test sample consists of a fiber of poly(methyl methacrylate) (PMMA) formed at the end of a small tapered glass capillary (Fig. 1b). The PMMA fiber terminates with an end radius of curvature below 1 μm. The capillary is glued to one prong of a commercial quartz tuning fork which oscillates at its mechanical resonance frequency. Precise measurements of the tuning fork frequency are used to monitor the approach of the polymer sample to the diamond surface, allowing for gentle contact of the polymer onto the surface. The NMR signal originating from the end of the polymer fiber is measured as a function of position as the fiber is slowly stepped across the diamond surface in a raster pattern. To avoid wear of the sample and to minimize contamination of the diamond surface, the sample is lifted off the diamond surface before each lateral scan step and then reapproached to the surface at each new position.

In order to achieve an acceptable signal-to-noise ratio (SNR) with tolerable scanning time, good coupling between the NV center and the sample protons is required. Since the proton dipole field falls inversely with the cube of the distance, the NV center should be located as close to the surface as possible while still maintaining adequate spin coherence. The NV centers used here were located 5 to 10 nm below the surface of a diamond substrate within a layer of isotopically pure carbon-12 diamond. Hahn spin echo measurements gave typical $T_2$ coherence times between 7 and 15 µs, extendable to more than 50 µs using multipulse decoupling techniques.

Several different measurement protocols have been shown to be effective in coupling the NV center to magnetic field signals from randomly polarized nuclear spins in the sample[14,15,19]. For this study, we use the method described by Staudacher et al.[14], where a sequence of periodic microwave pulses manipulates the precession of the NV center so as to accumulate net precession phase when the microwave pulse period is precisely half the period of the nuclear spin Larmor precession (Fig. 1c). To detect the presence of a nuclear spin signal, an optically detected multipulse spin echo experiment is performed and the resulting NV coherence $C(\tau)$ is measured as a function of the pulsing period $\tau$. For protons precessing in a static field $B_0$, the Larmor frequency is $f_L = (\gamma_n / 2\pi)B_0$ and a dip in NV coherence is expected for $\tau = \tau_L \equiv 1/2 f_L$. In our case, $\gamma_n / 2\pi = 42.56$ MHz/T, $B_0 = 38.6$ mT and $f_L = 1.64$ MHz.

As shown in Fig. 2, a clear dip in the coherence is seen at the expected value of $\tau_L = 304$ ns. It was verified that the dip position shifted as expected as a function of applied field. The sharpness of the dip is a consequence of the relatively narrow bandwidth of the proton precession (typically in the range of 10 – 50 kHz for organic solids at room temperature) and the frequency selectivity of the multipulse sequence (effective filter bandwidth on the order of $1/N\tau$, where $N$ is the number of $\pi$ pulses in the sequence). The coherence dip is seen even when the sample is retracted away from the diamond surface. This shows that the NV center is detecting the permanent presence of a thin proton layer coating the diamond surface, likely due to adsorbed water or hydrocarbons. Similar evidence for a ubiquitous proton-containing layer has been seen previously in MRFM experiments[5] and, more recently, in NV-based experiments[18]. An analysis of the coherence dips shows that the associated magnetic field amplitudes are approximately 330 nT-rms for the retracted case, and 485 nT-rms when the sample is in contact with the surface, assuming a Gaussian NMR linewidth of 20 kHz FWHM. These signals are consistent with an NV depth of 6.8 nm and a proton-containing layer 1.6 nm thick (see Supplementary Information).

## NMR imaging

During imaging measurements, decoherence due to the proton field must be distinguished from other sources of decoherence, such as randomly fluctuating magnetic fields from paramagnetic electron spins at the diamond surface or in the sample[24,25]. Thus, for reliable NMR discrimination, it is generally not sufficient to measure only the decoherence at the single value of $\tau = \tau_L$. Instead, we make additional measurements at $\tau = \tau_L \pm \Delta\tau$, where $\Delta\tau$ is chosen large enough to move the bandpass of the multipulse sequence to just outside of the NMR linewidth. In our case, we use $\Delta\tau = 32$ ns. The background decoherence is estimated by the average of these two measurements $C_0(\tau_L) = (1/2)\left[C(\tau_L + \Delta\tau) + C(\tau_L - \Delta\tau)\right]$. We define the NMR signal as a function of sample position to be $s(x,y) = 1 - C(\tau_L)/C_0(\tau_L)$. For a weakly coupled ensemble of randomly polarized nuclear spins, $s(x,y)$ will be limited to values between zero and one, with zero indicating no proton signal (Fig. S2). As a result of this limited dynamic range, it is important to choose an appropriate total evolution time $N\tau$ for the multipulse sequence. In our case, with NMR magnetic field amplitudes in the range of 300 nT-rms to 600 nT-rms, evolution times on the order of 30 μs are suitable.

Typical line scan results are shown in Fig. 3, where we plot the NMR signal vs. lateral position. As expected, a sharp stepwise increase in NMR signal is evident as the PMMA sample is scanned past the NV center. The NMR signal reaches a plateau value and is fairly constant when the sample is over the NV center. This plateau is to be expected for a relatively thick sample that deforms sufficiently to make a uniform Hertzian contact to the diamond surface. From the ratio of the feature step height to the standard deviation of the baseline, a signal-to-noise ratio of about 6 is estimated. The noise is primarily photon shot noise.

From the sharpness of the signal steps we estimate the spatial resolution to be approximately 12 nm. The resolution here was not set by the fundamentals of the NV response, but rather by technical issues. In particular, thermal drift and creep in the piezoelectric scan mechanism were the main limitations, exacerbated by the long duration of the scans. In Fig. 3, a data point was taken approximately every 4 minutes, with 3 minutes dedicated to photon counting and 1 minute for position control and the sample approach-retraction process. Consequently, the scanning process has the demanding requirement that nanometer accuracy be maintained over the many hours required to take a full image. Complete NMR images are shown in Fig. 4 for two different PMMA samples taken with two different NV centers. The height depicted in the images is proportional to $s(x, y)$.

## Point spread function (PSF)

Since there are no substantial magnetic field gradients present to affect the imaging response, the current imaging mode can be classified as a type of near field microscopy, where the resolution is set by the orientation and distance dependence of the proton-NV interaction[14]. In the weak signal limit ($s(x,y) \ll 1$), the PSF is proportional to the mean square field from the precessing protons along the axis of the NV center (see Supplemental Information). For our geometry with the (001)-oriented diamond surface and with the NV axis and applied field in the [111] direction, the mean-square oscillating field at the NV center from a proton at coordinate $(x, y, z)$ is given by

$$B_{rms}^2(x,y,z) = 2\left(\frac{\mu_0}{4\pi}\right)^2 \mu_n^2 \frac{(x+y+z+d)^2}{\left(x^2+y^2+(z+d)^2\right)^5}\left(x^2+y^2+(z+d)^2 - xy - (x+y)(z+d)\right),$$

where the NV center is assumed to be located at position (0,0,-*d*) with respect to the diamond surface and $\mu_n = \hbar\gamma_n/2$ is the magnetic moment of the proton. The actual response of the NV to this field will depend on the NMR line width and the filter function of the multipulse sequence.[14]

We plot in Fig. 5 the PSF based on the above equation for the case of a 10 nm deep NV center. The PSF falls to half its peak value at a distance of 1.3 nm above the diamond surface. The lateral resolution is somewhat asymmetrical as a result of the tilted [111]-orientation of the NV center. The lateral resolution is best in the [110] direction where the PSF width is 6 nm FWHM. Improved scan accuracy should allow this predicted resolution to be reached.

## Discussion

This work shows that two-dimensional nanoscale MRI can be achieved using the simple concept of scanning an organic sample past a near-surface NV center. This should be considered only a first step since there is much room for improvement in the technique. Since the noise is dominated by photon shot noise, improved photon collection efficiency using established photonic structures, such a solid immersion lenses[26] or microfabricated pillar waveguides[27], are obvious methods that would increase the rate of data collection. In addition, the coupling of the NV center to the sample protons can be increased by using NV centers located closer to the surface or by using improved detection protocols, such as double quantum magnetometry[28]. The imaging can, in principle, be extended to three dimensions with greatly enhanced spatial resolution by introducing sufficiently large magnetic gradients using small ferromagnets[23,29], nanoscale electromagnets[30] or, possibly, the gradient from the NV center itself.[16]

A number of issues were uncovered that need to be addressed if nanoscale MRI is to become useful for applications, such as molecular structure imaging. For example, the ubiquitous proton contamination layer is undesirable since it represents an interfering NMR signal source. Although spatial discrimination against this contamination layer might be achieved using field gradients, a better solution would be to find a way to eliminate the layer through environmental control. Another significant issue was discovered while trying to repeat scans using the same NV center. Some increase in the proton background signal and reduction of the NV $T_2$ was found after multiple scanning runs, necessitating the move to a fresh NV center. This may indicate that some contamination of the diamond surface occurs during the scan. In the future, it would be desirable for the scans to be performed in a true non-contact fashion, as has been developed for some types of atomic force microscopy[31].

## Methods

**Optical apparatus.** This study was performed at room temperature using a custom-built confocal microscope with fluorescence detection similar to that described previously[15]. Briefly, the negatively charged NV center was excited with 532 nm laser light and the resulting broadband red fluorescence was detected using appropriate wavelength filtering and single photon counting electronics. Illumination of the NV center was through the diamond substrate using a microscope objective designed for air-incident coverglass correction (Olympus UPLSAPO40X2, NA=0.95). Optical modeling showed that by using a 160 μm thick diamond substrate and setting the coverglass correction to 110 μm, acceptable compensation for spherical aberration induced by the diamond substrate could be obtained, despite the high refractive index of diamond. Typical optical spot size was below 500 nm FWHM. Laser power incident on the objective lens was approximately 300 μW with typical fluorescence photon counting rates of about 80 kHz (about half the saturation counting rate).

**Scanning.** Mechanical positioning of the sample was accomplished using commercial piezoelectric devices (Attocube models ANPx51 and ANPz51). Due to the long duration of the scans, significant challenges were encountered due to creep and thermal drift of the sample position with respect to the NV location. To improve thermal stability, the apparatus was placed in a box with its interior temperature controlled to within 0.1°C. In addition, the actual position of the sample with respect to the NV center was monitored during the scan by periodically taking reflection images of the diamond surface with the PMMA sample in contact with the diamond surface. The PMMA contact point appeared as a prominent dark dot in the image. The images were then cross-correlated with the reflection image taken at the start of the scan and with fluorescence images of the NV center taken periodically during the scan. In this way, the position of the sample with respect to the NV center could be determined with a precision of about 10 nm. The measured position information was used for the data shown in Figs. 3 and 4. In the case of Fig. 4b, the position information was used to help guide the scan into a more perfect raster pattern. The images in Fig. 4 were rendered using the WSXM presentation software[32] (www.nanotec.es), which required the data to be formatted onto a rectangular 2D grid. Since the measured data was on an irregular grid due to the drift and creep, a bilinear interpolation technique was used to create the regularized data array from the measured data.

**NV preparation.** A 50 nm thick epitaxial layer of isotopically pure carbon-12 diamond was grown by plasma-enhanced chemical vapor deposition[33] onto the (001) surface of a diamond substrate (Element Six) polished to a thickness of 160 μm (Syntek Co. LTD). Nitrogen-15 was ion implanted at an energy of 2.5 keV and then annealed for six hours at 850°C in vacuum ($1\times10^{-9}$ Torr). To remove graphitic contamination and to oxygen terminate the surface, the diamond was cleaned in a 200°C acid mixture (equal parts nitric, sulfuric and perchloric acids) for four hours and then heated to 425°C in pure oxygen for two hours. Gold microwires to generate the microwave magnetic field were fabricated on the diamond substrate by a lift-off technique using e-beam lithography and e-beam metal deposition.

**PMMA sample preparation.** Poly(methyl methacrylate) (Sigma Aldrich, MW ~ 996,000) was purified by dissolution in methylene chloride followed by filtration through a 0.2 μm PTFE membrane filter and precipitation into methanol. The precipitate was dried in a vacuum oven at 130°C for 12 h. The PMMA fiber was formed using a 12% solution in propylene glycol monomethyl ether acetate (PGMEA). The fiber was pulled from the blunted end of a 1 mm long borosilicate glass capillary drawn down to a 4 μm diameter tapered end. The glass capillary was attached to a quartz tuning fork using glue consisting of 30% PMMA (MW ~ 15,000) in diethyl phthalate. To apply the PMMA, the end of a 4 mm diameter glass rod was coated with the PMMA/PGMEA solution and mounted in the fixture of a micromanipulator. During the brief working time of the PMMA/PGMEA solution, the applicator rod was repeatedly brought into contact with and quickly pulled away from the end of the glass capillary until an appropriately shaped fiber was formed. The end of the fiber was then trimmed to length using PGMEA solvent dispensed through a fine capillary attached to the micromanipulator. Finally, the fiber-on-tuning fork was baked in a vacuum oven at 130° C for 12 h in 50 mTorr of flowing nitrogen in order to remove the solvent.

**Multipulse spin coherence measurements.** By alternating the sign of the initial π/2 pulse in the XY8-96 pulse sequence, the NV spin at the end of the sequence is projected onto either the $m_s = 0$ spin state or the $m_s = -1$ spin state, resulting in a difference in photon counting rate. The difference is maximum in the absence of decoherence and zero for full decoherence. To accumulate sufficient photon counts, the sequence was repeated at least 750k times for the two starting phases. The coherence $C(\tau)$ is given by the difference in the photon counts for the two projections normalized by the full available contrast in the absence of decoherence.

## Acknowledgements

We thank B. Myers and A. Jayich for helpful discussions. This work was supported by the DARPA QuASAR program, the Air Force Office of Scientific Research, the Center for Probing the Nanoscale at Stanford University (NSF grant PHY-0830228) and the IBM Corporation.

## Author contributions

D.R. and H.J.M. built the apparatus, performed the imaging experiments and analyzed the data. M.H.S. prepared the PMMA sample, annealed and acid cleaned the diamond, and provided temperature control. M.K. tested NV diamond samples and the scanning apparatus. K.O. and D.D.A. synthesized and characterized carbon-12 diamond layers. C.T.R. fabricated microwires. D.R. wrote the manuscript and incorporated comments from all authors.

## Author information

The authors declare no competing financial interests. Correspondence and requests for materials should be addressed to D.R. (rugar@us.ibm.com).

# References

1. Sidles, J. A. *et al.* Magnetic resonance force microscopy. *Rev. Mod. Phys.* **67,** 249 (1995).

2. Glover, P. & Mansfield, S. P. Limits to magnetic resonance microscopy. *Rep. Prog. Phys.* **65,** 1489–1511 (2002).

3. Poggio, M. & Degen, C. L. Force-detected nuclear magnetic resonance: recent advances and future challenges. *Nanotechnology* **21,** 342001 (2010).

4. Kuehn, S., Hickman, S. A. & Marohn, J. A. Advances in mechanical detection of magnetic resonance. *J. Chem. Phys.* **128,** 052208 (2008).

5. Degen, C. L., Poggio, M., Mamin, H. J., Rettner, C. T. & Rugar, D. Nanoscale magnetic resonance imaging. *Proc. Nat. Acad. Sci. USA* **106,** 1313–7 (2009).

6. Degen, C. L. Scanning magnetic field microscope with a diamond single-spin sensor. *Appl. Phys. Lett.* **92,** 243111 (2008).

7. Maze, J. R. *et al.* Nanoscale magnetic sensing with an individual electronic spin in diamond. *Nature* **455,** 644–7 (2008).

8. Balasubramanian, G. *et al.* Nanoscale imaging magnetometry with diamond spins under ambient conditions. *Nature* **455,** 648–51 (2008).

9. Taylor, J. M. *et al.* High-sensitivity diamond magnetometer with nanoscale resolution. *Nat. Phys.* **4,** 810–816 (2008).

10. Dobrovitski, V. V., Fuchs, G. D., Falk, A. L., Santori, C. & Awschalom, D. D. Quantum Control over Single Spins in Diamond. *Ann. Rev. Condens. Matter Phys.* **4,** 23–50 (2013).

11. Grotz, B. *et al.* Sensing external spins with nitrogen-vacancy diamond. *New J. Phys.* **13,** 055004 (2011).

12. Mamin, H. J., Sherwood, M. H. & Rugar, D. Detecting external electrons spins using nitrogen-vacancy centers. *Phys. Rev. B* **86,** 195422 (2012).

13. Jelezko, F. *et al.* Observation of coherent oscillation of a single nuclear spin and realization of a two-qubit conditional quantum gate. *Phys. Rev. Lett.* **93,** 130501 (2004).

14. Staudacher, T. *et al.* Nuclear magnetic resonance spectroscopy on a (5-nanometer)$^3$ sample volume. *Science* **339,** 561–3 (2013).

15. Mamin, H. J. *et al.* Nanoscale nuclear magnetic resonance with a nitrogen-vacancy spin sensor. *Science* **339,** 557–60 (2013).

16. Zhao, N., Honert, J., Schmid, B., Isoya, J. & Markham, M. Sensing remote nuclear spins. *Nat. Nanotech.* **7,** 657–662 (2012).

17. Kolkowitz, S., Unterreithmeier, Q., Bennett, S. D. & Lukin, M. D. Sensing distant nuclear spins with a single electron spin via dynamical decoupling. *Phys. Rev. Lett.* **109,** 137601 (2012).

18. Loretz, M., Pezzagna, S., Meijer, J. & Degen, C. L. Nanoscale nuclear magnetic resonance with a 1.9-nm-deep nitrogen-vacancy sensor. *Appl. Phys. Lett.* **104,** 033102 (2014).

19. Ohashi, K. *et al.* Negatively charged nitrogen-vacancy centers in a 5 nm thin C-12 diamond film. *Nano Lett.* **13,** 4733–4738 (2013).

20. Fuchs, G. D., Burkard, G., Klimov, P. V. & Awschalom, D. D. A quantum memory intrinsic to single nitrogen–vacancy centres in diamond. *Nat. Phys.* **7,** 789–793 (2011).

21. Pham, L. *et al.* Nanoscale NMR Spectroscopy and Imaging of Multiple Nuclear Species. *Bull. Am. Phys. Soc.* **59,** C6.00007 (2014).

22. Grinolds, M. S. *et al.* Nanoscale magnetic imaging of a single electron spin under ambient conditions. *Nat. Phys.* **9,** 215–219 (2013).

23. Grinolds, M. S. *et al.* Subnanometre resolution in three-dimensional magnetic resonance imaging of individual dark spins. *Nat. Nanotech.* **9,** 279–84 (2014).

24. Rosskopf, T. *et al.* Investigation of surface magnetic noise by shallow spins in diamond. *Phys. Rev. Lett.* **112,** 147602 (2014).

25. Myers, B. A. *et al.* Probing surface noise with depth-calibrated spins in diamond. *Phys. Rev. Lett.* (submitted)

26. Siyushev, P. *et al.* Monolithic diamond optics for single photon detection. *Appl. Phys. Lett.* **97,** 241902 (2010).

27. Babinec, T. M. *et al.* A diamond nanowire single-photon source. *Nat. Nanotech.* **5,** 195–9 (2010).

28. Mamin, H. J. *et al.* Multipulse Double-Quantum Magnetometry With Near-Surface Nitrogen Vacancy Centers. *Phys. Rev. Lett.* (submitted)

29. Mamin, H. J., Rettner, C. T., Sherwood, M. H., Gao, L. & Rugar, D. High field-gradient dysprosium tips for magnetic resonance force microscopy. *Appl. Phys. Lett.* **100,** 013102 (2012).

30. Nichol, J. M., Hemesath, E. R., Lauhon, L. J. & Budakian, R. Nanomechanical detection of nuclear magnetic resonance using a silicon nanowire oscillator. *Phys. Rev. B* **85,** 054414 (2012).

31. Giessibl, F. J. Advances in atomic force microscopy. *Rev. Mod. Phys.* **75,** 949–983 (2003).

32. Horcas, I. *et al.* WSXM: a software for scanning probe microscopy and a tool for nanotechnology. *Rev. Sci. Instrum.* **78,** 013705 (2007).

33. Ohno, K. *et al.* Engineering shallow spins in diamond with nitrogen delta-doping. *Appl. Phys. Lett.* **101,** 082413 (2012).

# Figure captions

**Figure 1 – Basic elements of the NMR imaging experiment.** (a) A PMMA polymer sample attached to a quartz tuning fork is brought into contact with a diamond substrate containing a near-surface NV center. The precessing protons in the sample are detected by the NV center as the sample is scanned past the NV center. The magnetic state of the NV center is read out optically via spin-dependent fluorescence. (b) Optical microscope image of the polymer sample. (c) The multipulse sequence manipulates the NV center so that its precession phase is selectively perturbed by the oscillating proton magnetic field when the pulse periodicity matches one-half the period of the proton field oscillations.

**Figure 2 – NV coherence in the presence of nearby protons.** A clear dip in coherence is observed when τ matches half the proton precession period. The decoherence is observed even when the polymer sample is retracted away from the NV center by several micrometers, indicating the permanent presence of a proton contamination layer on the surface of the diamond, possibly adsorbed water or hydrocarbons. Analysis of the coherence dips suggests that the NV is roughly 6.8 nm deep and the contamination layer is approximately 1.6 nm thick.

**Figure 3 – Line scans showing the NMR signal as a function of position.** A stepwise increase in NMR signal is seen when the sample is positioned over the NV center. The step height corresponds to a proton field of approximately $(300 \text{ nT})^2$. Spatial resolution is about 12 nm, limited primarily by the scan step size and scan position irregularities. For clarity, successive scans are displaced vertically by two divisions.

**Figure 4 – Two dimensional NMR images.** Results are from two different PMMA samples with two different NV centers. Height shown is proportional to $s(x,y)$. Apparent roughness is primarily due to photon shot noise. The image in (a) incorporates the line scan data shown in Fig. 3.

**Figure 5 – Calculated point spread function for a 10 nm deep NV.** (a) Vertical cross-section of the response shows that the signal falls by one-half for protons located 1.3 nm above the diamond surface. The maximum response is slightly shifted laterally with respect to the NV position due to the [111] orientation of the NV axis. (b) Lateral response at the diamond surface

is shown. The PSF width is slightly asymmetrical, with the narrowest being 6 nm FWHM in the [110] direction.

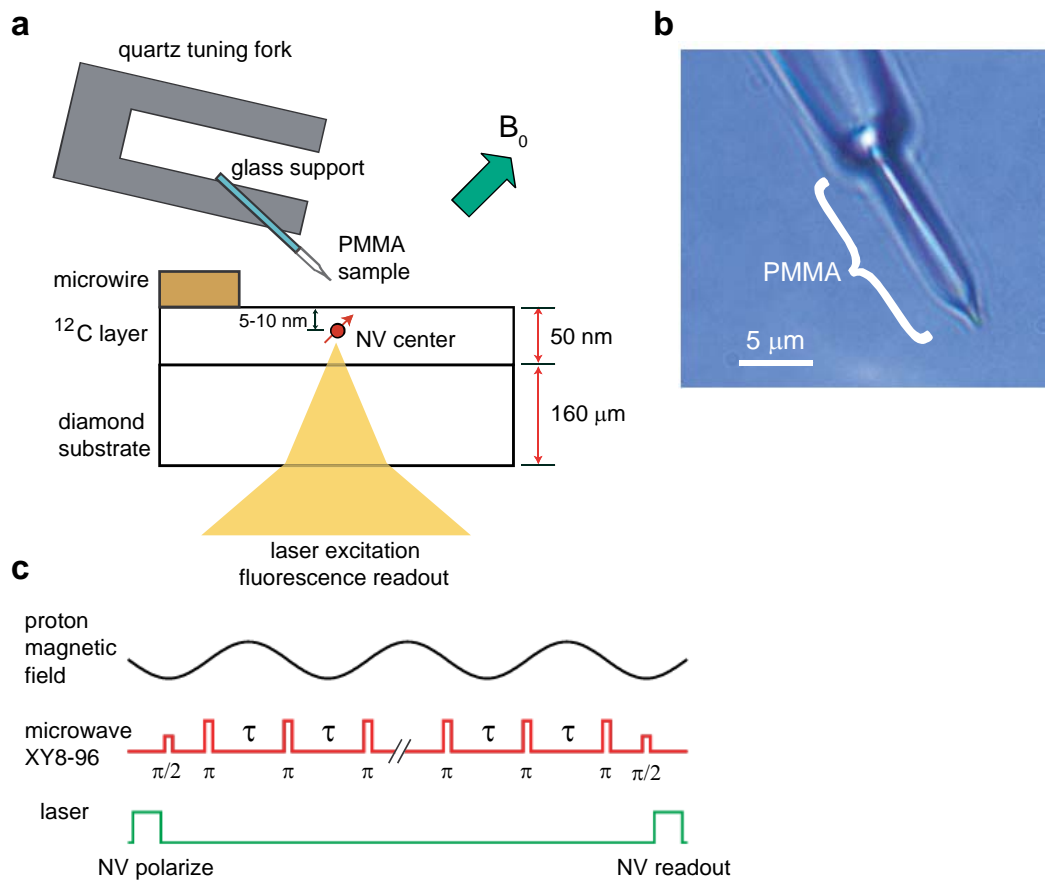

**Figure 1 – Basic elements of the NMR imaging experiment.** (a) A PMMA polymer sample attached to a quartz tuning fork is brought into contact with a diamond substrate containing a near-surface NV center. The precessing protons in the sample are detected by the NV center as the sample is scanned past the NV center. The magnetic state of the NV center is read out optically via spin-dependent fluorescence. (b) Optical microscope image of the polymer sample. (c) The multipulse sequence manipulates the NV center so that its precession phase is selectively perturbed by the oscillating proton magnetic field when the pulse periodicity matches one-half the period of the proton field oscillations.

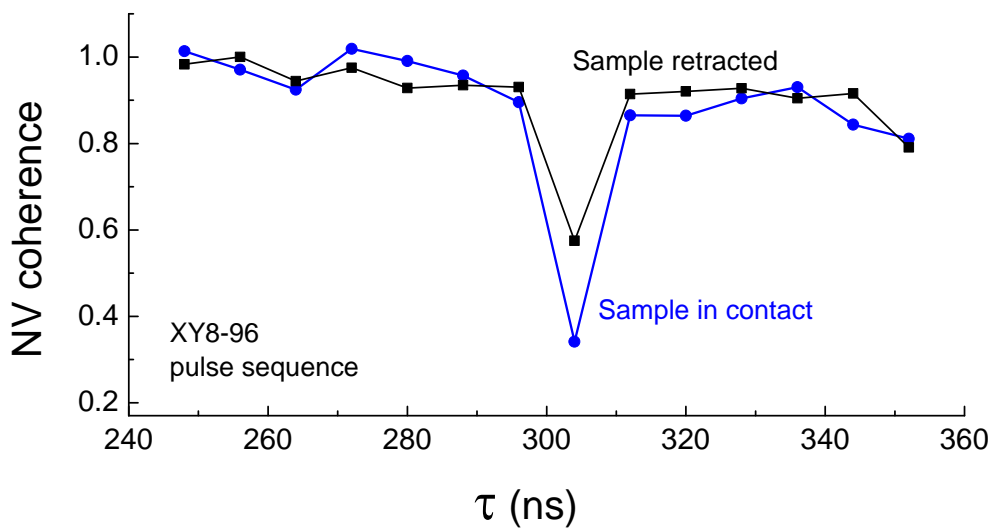

**Figure 2 – NV coherence in the presence of nearby protons.** A clear dip in coherence is observed when τ matches half the proton precession period. The decoherence is observed even when the polymer sample is retracted away from the NV center by several micrometers, indicating the permanent presence of a proton contamination layer on the surface of the diamond, possibly adsorbed water or hydrocarbons. Analysis of the coherence dips suggests that the NV is roughly 6.8 nm deep and the contamination layer is approximately 1.6 nm thick.

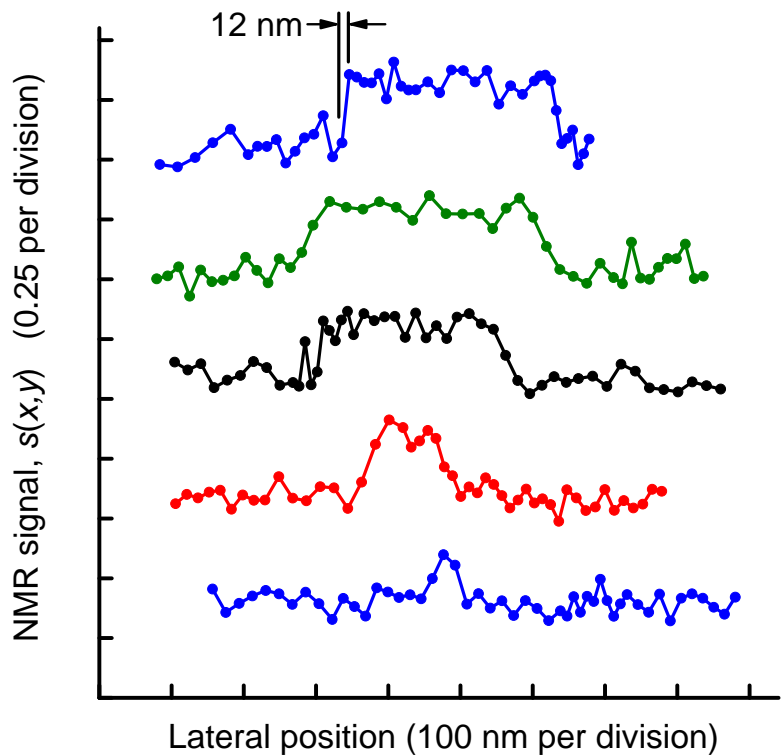

**Figure 3 – Line scans showing the NMR signal as a function of position.** A stepwise increase in NMR signal is seen when the sample is positioned over the NV center. The step height corresponds to a proton field of approximately $(300 \text{ nT})^2$. Spatial resolution is about 12 nm, limited primarily by the scan step size and scan position irregularities. For clarity, successive scans are displaced vertically by two divisions.

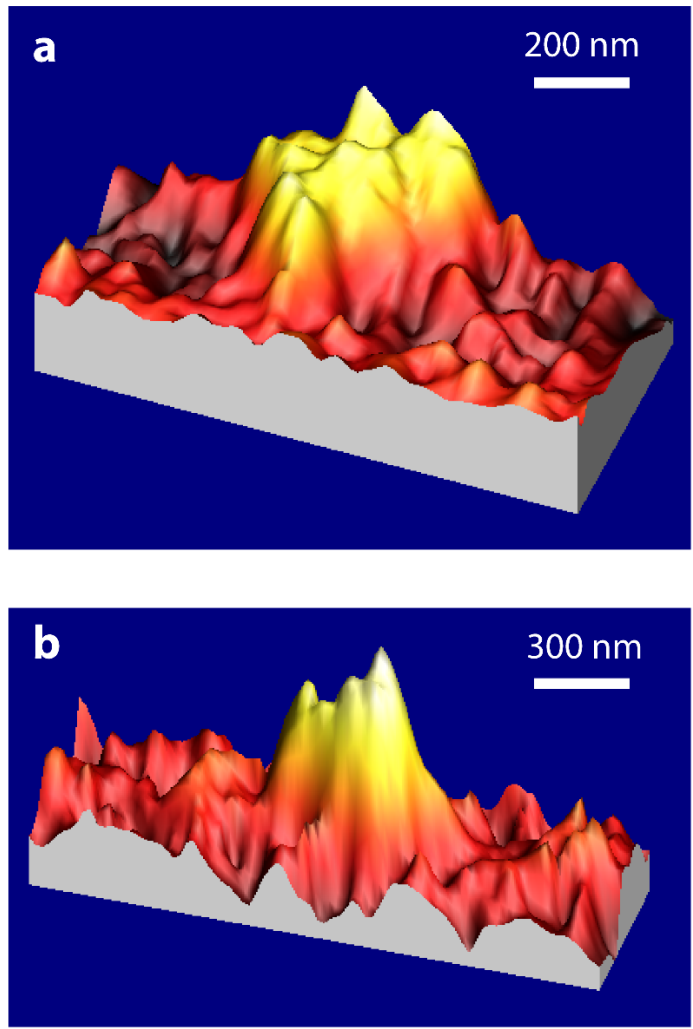

**Figure 4 – Two dimensional NMR images.** Results are from two different PMMA samples with two different NV centers. Height shown is proportional to $s(x,y)$. Apparent roughness is primarily due to photon shot noise. The image in (a) incorporates the line scan data shown in Fig. 3.

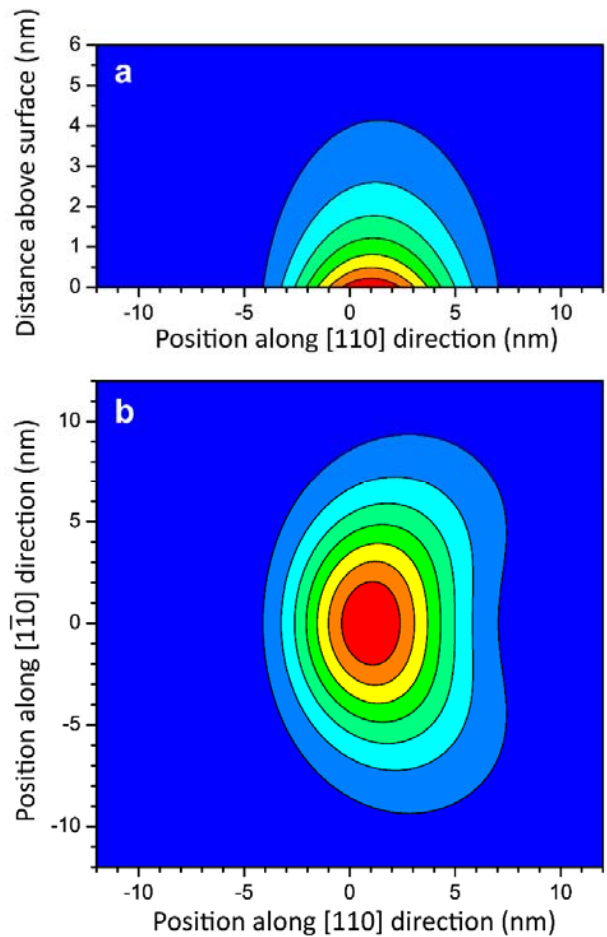

**Figure 5 – Calculated point spread function for a 10 nm deep NV.** (a) Vertical cross-section of the response shows that the signal falls by one-half for protons located 1.3 nm above the diamond surface. The maximum response is slightly shifted laterally with respect to the NV position due to the [111] orientation of the NV axis. (b) Lateral response at the diamond surface is shown. The PSF width is slightly asymmetrical, with the narrowest being 6 nm FWHM in the [110] direction.

# Supplementary Information

**Proton magnetic resonance imaging with a nitrogen-vacancy spin sensor**

D. Rugar[1]\*, H. J. Mamin[1], M. H. Sherwood[1], M. Kim[1,2], C. T. Rettner[1], K. Ohno[3] and D. D. Awschalom[3,4]

[1]IBM Research Division, Almaden Research Center, San Jose, CA 95120, USA

[2]Center for Probing the Nanoscale, Stanford University, Stanford, CA 94305, USA

[3]Center for Spintronics and Quantum Computation, University of California, Santa Barbara, CA 93106, USA

[4]Institute for Molecular Engineering, University of Chicago, IL 60637, USA

\*For correspondence: rugar@us.ibm.com

6/8/2014 12:54 PM

## NV spin coherence in presence of random magnetic noise

For a multipulse experiment of the type considered in this paper, the loss of spin coherence in the presence of Gaussian random magnetic noise can be expressed as[1,2] $C(\tau) = \exp[-\chi(\tau, N)]$, where $N$ is the number of pi pulses in the sequence, $\tau$ is the pulse spacing and

$$\chi(\tau, N) = \frac{\gamma_{NV}^2}{\pi} \int_0^\infty d\omega \, S_{z'}(\omega) \frac{F(\omega; \tau, N)}{\omega^2} \,. \tag{S1}$$

Here, $\gamma_{NV} = 2\pi \times 28$ GHz/T is the gyromagnetic ratio for the NV center, $S_{z'}(\omega)$ is the double-sided power spectral density of the magnetic noise component along the [111] axis of the NV center (denoted as the $z'$ axis) and $F(\omega; \tau, N)$ is a dimensionless "filter function" that specifies the bandpass of the multipulse sequence. For the XY8-N sequence used in this paper (or similar CPMG-like sequences), the filter function for even values of $N$ is[1]

$$F(\omega; \tau, N) = 8 \sin^2(\omega N \tau / 2) \frac{\sin^4(\omega \tau / 4)}{\cos^2(\omega \tau / 2)} \tag{S2}$$

The filter function for our XY8 sequence with $N$=96 and $\tau$ =304 ns is plotted in Fig. S1. The full width at half maximum (FWHM) is 30 kHz, or approximately $1/N\tau$.

## NV signal response to proton magnetic field

The spectral density of the random magnetic field along the NV axis can be considered as the sum of two parts:

$$S_{z'}(\omega) = S_n(\omega) + S_{bg}(\omega) \,, \tag{S3}$$

where $S_n(\omega)$ is the contribution from the $^1$H nuclear spin precession and $S_{bg}(\omega)$ represents all other background field contributions that lead to NV spin decoherence. We assume the precessing component has a Gaussian line shape with double-sided spectral density of the form

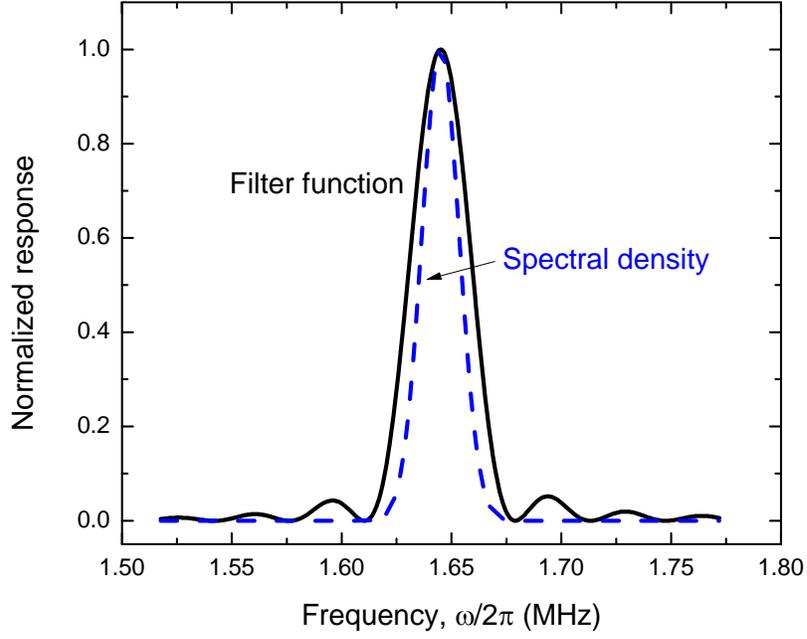

**Figure S1** – Normalized plots of the filter function and proton magnetic field spectral density. Filter function assumes $N = 96$ and $\tau = 304$ ns. Filter bandpass is 30 kHz FWHM. The magnetic field spectral density is assumed to have a Gaussian line shape centered at 1.64 MHz with FWHM of 20 kHz.

$$S_n(\omega) = B_{rms}^2 \left(\pi / 2\sigma^2\right)^{1/2} \left\{ \exp\left[-(\omega - \omega_L)^2 / 2\sigma^2\right] + \exp\left[-(\omega + \omega_L)^2 / 2\sigma^2\right] \right\}, \tag{S4}$$

where the Gaussian width parameter $\sigma$ is related to $\Delta f$, the FWHM linewidth in Hz, by $\sigma = \pi \Delta f / \sqrt{2 \ln 2} \approx 2.67 \Delta f$. Note that this spectral density has the normalization

$$\frac{1}{2\pi} \int_{-\infty}^{\infty} S_n(\omega) \, d\omega = B_{rms}^2. \tag{S5}$$

Combining (S1) and (S3), we can write

$$\chi(\tau, N) = \chi_n(\tau, N) + \chi_{bg}(\tau, N), \tag{S6}$$

where

$$\chi_n(\tau, N) = \frac{\gamma_{NV}^2}{\pi} \int_0^\infty d\omega \, S_n(\omega) \frac{F(\omega; \tau, N)}{\omega^2} \tag{S7}$$

and

$$\chi_{bg}(\tau, N) = \frac{\gamma_{NV}^2}{\pi} \int_0^\infty d\omega \, S_{bg}(\omega) \frac{F(\omega; \tau, N)}{\omega^2} \, . \tag{S8}$$

The NMR signal, defined in the main text as $s = 1 - C(\tau_L)/C_0(\tau_L)$, can then be written as

$$s = 1 - \frac{\exp\left[-\left[\chi_n(\tau_L, N) + \chi_{bg}(\tau_L, N)\right]\right]}{\exp\left(-\chi_{bg}(\tau_L, N)\right)} = 1 - \exp\left(-\chi_n(\tau_L, N)\right) \tag{S9}$$

In the limit of a weak proton signal where $\chi_n$ is small, the signal is proportional to $B_{rms}^2$:

$$s \approx \chi_n(\tau_L, N) = \varepsilon B_{rms}^2 \, , \tag{S10}$$

where $\varepsilon$ characterizes the overlap between the filter function and the proton line shape

$$\varepsilon = \frac{\gamma_{NV}^2}{\pi} \int_0^\infty d\omega \, g(\omega) \frac{F(\omega; \tau_L, N)}{\omega^2} \, , \tag{S11}$$

and $g(\omega)$ is the line shape function (for positive frequencies only)

$$g(\omega) = \left(\pi / 2\sigma^2\right)^{1/2} \exp\left[-(\omega - \omega_L)^2 / 2\sigma^2\right] \, . \tag{S12}$$

Combining equations (S2), (S4), (S7) and (S9), we plot in Fig. S2 the signal $s$ vs. $B_{rms}$ for parameters relevant to our experiment: $N = 96$, $\tau = 304$ ns, $f_L = 1.645$ MHz and $\Delta f = 20$ kHz.

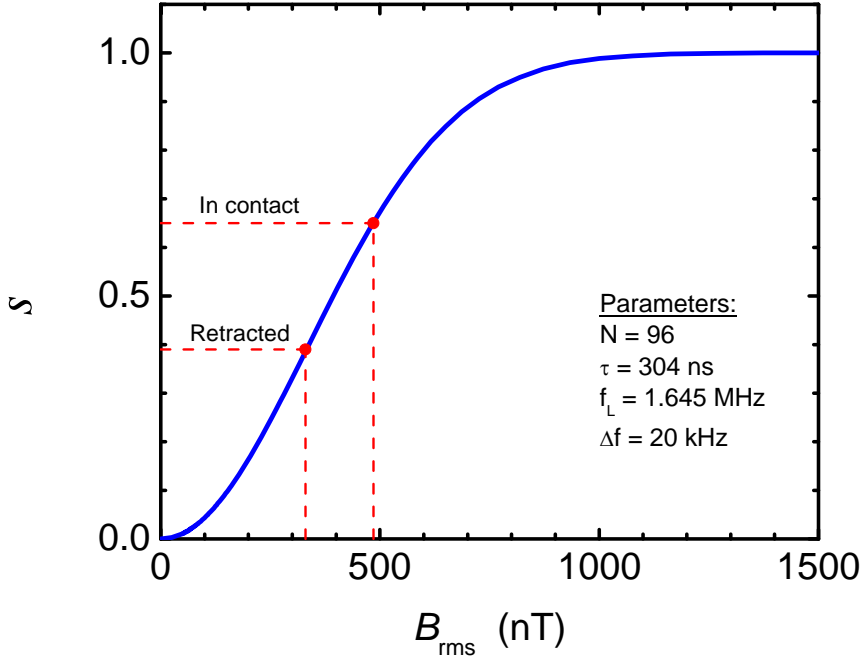

**Figure S2** – NV signal response *s* as a function of the rms amplitude of the magnetic field. Signal values and associated proton magnetic fields for the coherence dips in Fig. 2 (main text) are indicated.

## Proton field calculations and the point spread function

The mean-square field $B_{rms}^2$ originates from the protons that precess about the static field, which is oriented along the [111] crystal axis (defined here as the $z'$ axis). For a narrow-band filter function centered at the proton Larmor frequency, the most important contributions to the signal come from the perpendicular components of the proton moment ($m_{x'}$ and $m_{y'}$) which produce an oscillating field component along the $z'$ axis of the NV center. This field for a NV center located at position $(0,0,-d)$ from a proton located at position $(x,y,z)$ is given by

$$\mathbf{B}\cdot\hat{\mathbf{z}}' = \frac{3\mu_0}{4\pi r^5}\left[\left(\hat{\mathbf{x}}'m_{x'}(t)+\hat{\mathbf{y}}'m_{y'}(t)\right)\cdot\mathbf{r}\right]\left(\mathbf{r}\cdot\hat{\mathbf{z}}'\right) , \quad\quad (S13)$$

where $\hat{\mathbf{x}}'$, $\hat{\mathbf{y}}'$ and $\hat{\mathbf{z}}'$ constitute an orthonormal basis with $\hat{\mathbf{z}}' = (1/\sqrt{3})(\hat{\mathbf{x}} + \hat{\mathbf{y}} + \hat{\mathbf{z}})$ aligned in the [111] direction and $\mathbf{r} = -(x\hat{\mathbf{x}} + y\hat{\mathbf{y}} + (z+d)\hat{\mathbf{z}})$ is the vector from the proton to the NV center. For random, unpolarized protons, the mean square field is[3]

$$B_{rms}^2 \equiv \langle (\mathbf{B} \cdot \hat{\mathbf{z}}')^2 \rangle = \left(\frac{\mu_0}{4\pi}\right)^2 \frac{9(\mathbf{r} \cdot \hat{\mathbf{z}}')^2}{r^{10}} \left[ \langle m_x^2 \rangle (\hat{\mathbf{x}}' \cdot \mathbf{r})^2 + \langle m_y^2 \rangle (\hat{\mathbf{y}}' \cdot \mathbf{r})^2 \right]. \tag{S14}$$

To determine the appropriate values for $\langle m_{x'}^2 \rangle$ and $\langle m_{y'}^2 \rangle$, we note that the expectation value for $\langle m^2 \rangle$ from quantum mechanics is $\langle m^2 \rangle = \hbar^2 \gamma_n^2 \langle \tilde{I}^2 \rangle = \hbar^2 \gamma_n^2 I(I+1)$, where $\tilde{I}$ is the spin angular momentum operator. Since $I = 1/2$ for a proton, we obtain $\langle m^2 \rangle = \frac{3}{4} \hbar^2 \gamma_n^2$. For unpolarized nuclei, each orthogonal component will take on 1/3 of this value, or $\langle m_{x'}^2 \rangle = \langle m_{y'}^2 \rangle = \frac{1}{3} \langle m^2 \rangle = \hbar^2 \gamma_n^2 / 4$. With this result and after some additional algebra, we obtain the expression in the main text

$$B_{rms}^2(x,y,z) = 2\left(\frac{\mu_0}{4\pi}\right)^2 \mu_n^2 \frac{(x+y+z+d)^2}{(x^2+y^2+(z+d)^2)^5} \left(x^2 + y^2 + (z+d)^2 - xy - (x+y)(z+d)\right),$$

(S15)

where for a proton $\mu_n = \hbar \gamma_n / 2 = 1.41 \times 10^{-26}$ J/T.

Since, according to (S10), the signal is proportional to $B_{rms}^2$ for small signals, (S15) gives the basic form of the point spread function.

## Field from a proton containing layer

The mean square field for a proton-containing layer on the surface of the diamond can be found by integration of equation (S15):

$$B_{rms}^2 = \rho_N \int_0^h dz \int_{-\infty}^{\infty} dy \int_{-\infty}^{\infty} dx\, B_{rms}^2(x,y,z)$$

$$= \frac{5\pi}{24}\left(\frac{\mu_0}{4\pi}\right)^2 \mu_n^2 \rho_N \left[\frac{1}{d^3} - \frac{1}{(d+h)^3}\right]$$

(S16)

where $h$ is the film thickness and $\rho_N$ is the proton number density. This expression agrees with the result given by Loretz et al.[4], who considered the particular case of an infinitely thick film. Figure S3 shows plots of the field at the NV center as a function of NV depth for a 1.6 nm thick layer and an infinitely thick layer of protons, assuming a proton density representative of PMMA: $\rho_N = 5.7 \times 10^{28}$ m$^{-3}$ (57 protons per nm$^3$).

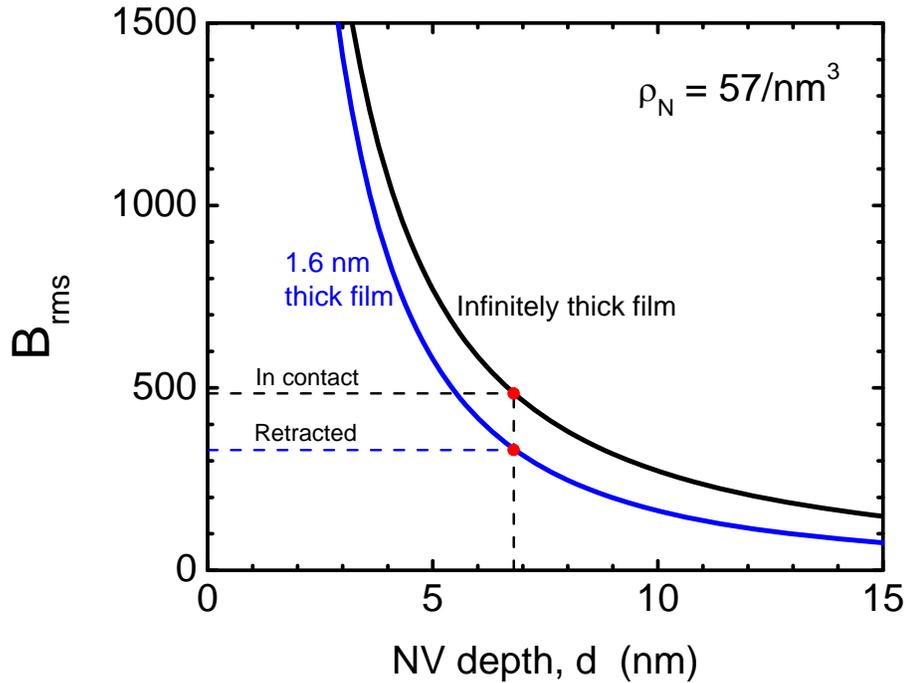

**Figure S3** – Theoretical magnetic field at the NV center from protons on the diamond surface as a function of NV depth. The two field values corresponding to the coherence dips in Fig. 2 (main text) are indicated.

## Analysis of coherence dips in Fig. 2

The coherence dips in Fig. 2 of the main text correspond to signal values $s$ of 0.65 for the contact case and 0.39 for the retracted case. Using equations (S2), (S4), (S7) and (S9), we can find the magnetic field $B_{rms}^2$ responsible for these signal values.. We assume the proton linewidth $\Delta f = 20$ kHz based on previous work[5] and take the experimental parameters, $N = 96$, $\tau = 304$ ns and $\omega_L = 2\pi \times 1.64$ MHz. We find $B_{rms}^2 = (485 \text{ nT})^2$ when in contact and $(330 \text{ nT})^2$ when the sample is retracted. These field amplitudes can be read directly from the plot in Fig. S2.

Based on these values of $B_{rms}^2$, equation (S16) can be inverted to learn about the NV depth and the thickness of the permanent proton-containing layer. Since the sample is relatively large compared to the sensitive volume of the point spread function, we can use (S16) with $h = \infty$ for the case where the sample is in contact with the diamond. Assuming a proton density of $\rho_N = 5.7 \times 10^{28}$ m$^{-3}$, we find that the field $B_{rms}^2 = (485 \text{ nT})^2$ for the contact case implies a NV depth of $d = 6.8$ nm (Fig. S3). Next we consider the retracted case where $B_{rms}^2 = (330 \text{ nT})^2$. Taking 6.8 nm as the NV depth and assuming that the permanent proton-containing layer (adsorbed water or hydrocarbons) has proton density similar to PMMA, we use (S16) to find the layer thickness to be 1.6 nm.

# References

1. Cywiński, Ł., Lutchyn, R., Nave, C. & Das Sarma, S. How to enhance dephasing time in superconducting qubits. *Phys. Rev. B* **77,** 174509 (2008).

2. Bar-Gill, N. *et al.* Suppression of spin-bath dynamics for improved coherence of multi-spin-qubit systems. *Nat. Commun.* **3,** 858 (2012).

3. Staudacher, T. *et al.* Nuclear magnetic resonance spectroscopy on a (5-nanometer)$^3$ sample volume. *Science* **339,** 561–3 (2013). See supplement for a derivation similar to that presented here, except for a key difference in the values chosen for $<m_x^2>$ and $<m_y^2>$.

4. Loretz, M., Pezzagna, S., Meijer, J. & Degen, C. L. Nanoscale nuclear magnetic resonance with a 1.9-nm-deep nitrogen-vacancy sensor. *Appl. Phys. Lett.* **104,** 033102 (2014).

5. Mamin, H. J. *et al.* Nanoscale nuclear magnetic resonance with a nitrogen-vacancy spin sensor. *Science* **339,** 557–60 (2013).